\documentclass{llncs}

\usepackage[utf8]{inputenc}
\usepackage[T1]{fontenc}
\usepackage{amsmath}
\usepackage{amssymb}

\usepackage{times}

\usepackage{xcolor}
\definecolor{mygreen}{RGB}{0,159,57}
\definecolor{myyellow}{RGB}{255,200,0}
\definecolor{myorange}{RGB}{255,140,0}
\definecolor{myblue}{RGB}{0,50,255}
\colorlet{myred}{red!90}

\usepackage{fancyvrb}

\usepackage{tikz}
\usetikzlibrary{calc}

\usepackage{enumitem}  
\setlist{nolistsep} 

\usepackage{placeins}

\usepackage{bussproofs}

\newcommand{\satvis}{\textsc{SatVis}}
\newcommand{\vampire}{\textsc{Vampire}}

\usepackage{url}

\usepackage{rotating}
\usepackage{multirow}
\usepackage{graphicx}

\usepackage{times}
\begin{document}

\title{
  Interactive Visualization of Saturation Attempts in Vampire
}

\author{
  Bernhard Gleiss\inst{1}
  \and
  Laura Kov{\'a}cs\inst{1 \and 2}
  \and
  Lena Schnedlitz\inst{1}
}

\institute{
  TU Wien, Austria
  \and
  Chalmers University of Technology, Sweden
}

\maketitle        

\vspace*{-.5em}

\begin{abstract}

Many applications of formal methods require
automated reasoning about system properties, such as system safety and security.
To improve the performance of automated reasoning engines, such as SAT/SMT solvers and first-order theorem prover, it is necessary to understand both the successful and failing attempts of these engines towards producing formal certificates, such as logical proofs and/or models.
Such an analysis is challenging due to the large number of logical formulas generated during proof/model search. 
In this paper we focus on saturation-based first-order theorem proving and introduce the \satvis{} tool for interactively visualizing saturation-based proof attempts in first-order theorem proving. 
We build \satvis{} on top of the world-leading theorem prover \vampire{}, by interactively visualizing the saturation attempts of \vampire{} in \satvis{}.
Our work combines the automatic layout and visualization of the derivation graph induced by the saturation attempt with interactive transformations and search functionality.
As a result, we are able to analyze and debug (failed) proof attempts of \vampire{}. 
Thanks to its interactive visualisation, we believe \satvis{} helps both experts and non-experts in theorem proving to understand first-order proofs and analyze/refine failing proof attempts of first-order provers.
\end{abstract}\vspace*{-2em}

\section{Introduction}

%\bg{
%	Possible comments from the reviewers we should invalidate/refute:
%	\begin{itemize}
%		\item "Visualizing a derivation graph is straightforward"
%		\item "Visualizing a derivation graph can be done using Graphviz"
%		\item "It's not necessary to use an extra visualization tool, since any of the information is already contained in the text-based output.
%		\item "It's not necessary to use an extra visualization tool, one just needs to efficiently search through the text-based output using some regex-search tricks or using some script"
%		\item "This tool is Vampire-specific, but I would favor a tool with support for any solver which is able to produce derivations."
%		\item "Why do we need another \emph{proof}-visualization?"
%	\end{itemize}
%}
%~\\

Many applications of formal methods, such as program analysis and verification,
require automated reasoning about system properties, such as program safety, security and reliability.
Automated reasoners, such as SAT/SMT solvers~\cite{CVC4,Z3} and first-order theorem provers~\cite{kovacs2013first,ESchulz},
have therefore become a key backbone of rigorous system engineering.
For example, proving properties over the computer memory  relies on first-order reasoning with both  quantifiers and 
 integer
arithmetic. \\[-.75em]

Saturation-based theorem proving is the leading approach for automating reasoning in full first-order logic.
In a nutshell, this approach negates a given goal and saturates its given set of input formulas (including the negated goal),
by deriving logical consequences of the input using a logical inference system, such as binary resolution or superposition.
Whenever a contradiction (false) is derived, the saturation process terminates reporting validity of the input goal.
State-of-the-art theorem provers, such as \vampire{}~\cite{kovacs2013first} and E~\cite{ESchulz}, implement saturation-based proof search using the (ordered) superposition calculus~\cite{Sup01}.
These provers rely  on powerful indexing algorithms, selection functions and term orderings for making saturation-based theorem proving efficient and scalable to a large set of first-order formulas, as evidenced in the yearly CASC system competition of first-order provers~\cite{Sut07-CSR}. \\[-.75em]%practically useful.  and carefully chosen heuristics, resulting in the fastest tool currently available to reason in full first-order logic \cite{Sut07-CSR}.

Over the past years, saturation-based theorem proving has been extended
to first-order logic with theories, such as arithmetic, theory of arrays and
algebraic datatypes~\cite{kovacs2017coming}. Further, 
first-class boolean sorts and if-then-else and let-in constructs have also been introduced as extensions 
to the input syntax of first-order theorem provers~\cite{FOOL18}.  Thanks to these recent developments,
first-order theorem provers became better suited in applications of formal methods, being for example a competitive alternative to SMT-solvers \cite{CVC4,Z3} in  software verification and program analysis. Recent editions of the SMT-COMP\footnote{\url{https://smt-comp.github.io/}}
 and CASC system competitions show, for example, that \vampire{} successfully competes against the leading SMT solvers Z3~\cite{Z3} and CVC4~\cite{CVC4} and vice-versa. \\[-.75em]

 By leveraging the best practices in first-order theorem proving in combination with SMT solving,
 in our recent work~\cite{rapidArxiv} we showed that correctness of a software program can be reduced to a validity problem in first-order logic. We use \vampire{} to prove the resulting encodings, outperforming SMT solvers. Our  initial results demonstrate that first-order theorem proving is well-suited for  applications of (relational) verification, such as safety and non-interference. Yet, our results also show that the performance of the prover crucially depends on the
 logical representation of its input problem and the deployed reasoning strategies during proof search.
 As such, users and developers of first-order provers, and automated reasoners in general, typically face the burden of analysing (failed) proof attempts produced by the prover,
 with the ultimate goal to refine the input and/or proof strategies making the prover succeed in proving its input.
 Understanding (some of) the reasons why the prover failed is however very hard and requires a considerable
 amount of work by highly qualified experts in theorem proving,
 hindering thus the use of theorem provers in many application domains. \\[-.75em]

 In this paper we address this challenge and {\it introduce the \satvis{} tool to ease the task of 
 analysing failed proof attempts in saturation-based reasoning}. We designed \satvis{} to support
 interactive visualization of the saturation algorithm used in \vampire{},
 with the goal to ease the manual analysis of \vampire{} proofs as well as failed proof attempts in \vampire{}.
 Inputs to \satvis{} are proof (attempts) produced by \vampire{}.
 Our tool  consists of (i) an explicit visualization of the DAG-structure of the saturation proof (attempt) of \vampire{}
 and (ii) interactive transformations of the DAG for pruning and reformatting the proof (attempt).
%  As a result, \satvis{} produces a DAG visualisation of the \vampire{} proof (attempt).
 In its current setting, \satvis{} can be used only in the context of \vampire{}.
 Yet, by parsing/translating proofs (or proof attempts) of other provers into the \vampire{} proof format,
\satvis{} can be used in conjunction with other provers as well. \\[-.75em]

When feeding \vampire{} proofs to \satvis{}, \satvis{} supports both users and developers of \vampire{} to understand and refactor \vampire{} proofs, and to manually proof check soundness of \vampire{} proofs.
When using \satvis{} on failed proof attempts of \vampire{},
\satvis{} supports users and developers of \vampire{} to analyse how \vampire{} explored its search space during proof search, that is, to understand which clauses were derived and why certain clauses have not been derived at various steps during saturation. By doing so, the \satvis{} proof visualisation framework gives valuable insights on how to  
revise the input problem encoding of \vampire{} and/or implement domain-specific optimizations in \vampire{}. 
We therefore believe that \satvis{} improves the state-of-the-art in the use and applications of theorem proving at least in the following scenarios: (i) helping \vampire{} developers to debug and further improve \vampire{},
(ii) helping \vampire{} users to tune \vampire{} to their applications, by not treating \vampire{} as a black-box but by understanding and using its appropriate proof search options;
and 
(iii) helping unexperienced users in saturation-based theorem proving to learn using \vampire{} and first-order proving in general.

\paragraph{\bf Contributions.} The contribution of this paper comes with the design of the \satvis{} tool for analysing proofs, as well as proof attempts of the \vampire{} theorem prover.
\satvis{} is available at:\vspace*{-.5em}
\[
\text{\url{https://github.com/gleiss/saturation-visualization}. }\vspace*{-.5em}\]
We overview  proof search steps in \vampire{} specific to \satvis{} (Section~\ref{sec:search}), discuss the challenges we faced for analysing proof attempts of \vampire{} (Section~\ref{sec:analysis}), and describe implementation-level details of \satvis{} 1.0 (Section~\ref{sec:implementation}). \vspace*{-.5em}

%\todo{shift somewhere: Each of the saturation-based theorem provers E, Spass, Otter and Prover9 is able to output newly generated and activated clauses. It would therefore be easy to support %the visualization of saturation attempts for any such prover by extending our tool with a custom parser for the output generated by that prover.}

\paragraph{Related work.} While standardizing the input format of automated reasoners is an active research topic, see e.g. the SMT-LIB~\cite{barrett2017smtlib} and TPTP~\cite{Sut07-CSR} standards,
coming up with an input standard for representing and analysing proofs and proof attempts of automated reasoners has received so far very little attention. The TSTP library~\cite{Sut07-CSR} provides input/output standards for automated theorem proving systems. Yet, unlike \satvis{}, TSTP does not analyse proof attempts but only
supports the examination of first-order proofs. We note that \vampire{} proofs (and proof attempts) contain first-order formulas with theories, which is not fully supported by TSTP. 

Using a graph-layout framework, for instance Graphviz~\cite{Gansner00anopen}, it is relatively 
%Given enough information on the saturation attempt,
%it is
straightforward to visualize the DAG derivation graph induced by a saturation attempt of a first-order prover.
For example, the theorem prover E~\cite{ESchulz} is able to directly output its saturation attempt as an input file for Graphviz. 
The visualizations generated in this way are useful however only for analyzing small derivations with at most 100 inferences,
but cannot practically be used to analyse and manipulate larger proof attempts.
We note that it is quite common to have first-order proofs and proof attempts with more than 1,000 or even 10,000 inferences, especially in applications of theorem proving in software verification, see e.g.~\cite{rapidArxiv}.  In our \satvis{} framework,  the interactive features of our tool allow one to analyze such large(r) proof attempts.

The framework~\cite{z3-axiom-profiler} eases the manual analysis of proof attempts in Z3~\cite{Z3} by visualizing quantifier instantiations, case splits and conflicts. While both \cite{z3-axiom-profiler} and \satvis{} are built for analyzing (failed) proof attempts, they target different architectures (SMT-solving resp. superposition-based proving) and therefore differ in their input format and in the information they visualize.
The frameworks~\cite{BYRNES200923,DBLP:journals/corr/LibalRR14} visualize proofs derived  in a natural deduction/sequent calculus. Unlike these approaches, \satvis{} targets clausal derivations generated by saturation-based provers using the superposition inference system.
As a consequence, our tool can be used to focus only on the clauses that have been actively used during proof search,
instead of having to visualize the entire set of clauses, including unused clauses during proof search.
We finally note that
proof checkers, such as DRAT-trim~\cite{DRAT}, support the soundness analysis of each inference step of a proof, and do not focus on failing proof attempts nor do they visualize proofs. 

\vspace*{-.5em}

\section{Proof Search in \vampire}\label{sec:search}% / Preliminaries}
We first present the key ingredients for proof search in \vampire{}, relevant to analysing saturation attempts.

\paragraph{Derivations and proofs.}
An \emph{inference} $I$ is a tuple $(F_1,\dots,F_n,F)$, where $F_1,\dots,F_n,F$ are formulas. The formulas $F_1,\dots,F_n$ are called the \emph{premises} of $I$ and $F$ is called the \emph{conclusion} of $I$. In our setting, an \emph{inference system} is a set of inferences and we rely on the superposition inference systems~\cite{Sup01}.
An axiom of an inference system is any inference with $n=0$. 
Given an inference system $\mathcal{I}$, a \emph{derivation} from axioms $A$ is an acyclic directed graph (DAG), where (i) each node is a formula and (ii) each node either is an axiom in $A$ and does not have any incoming edges, or is a formula $F \notin A$, such that the incoming edges of $F$ are exactly $(F_1,F),\dots,(F_n,F)$ and there exists an inference $(F_1,\dots,F_n,F) \in \mathcal{I}$. A refutation of axioms $A$ is a derivation which contains the empty clause $\bot$ as a node. A derivation of a formula $F$ is
called a {\it proof} of $F$ if it is finite and all leaves in the derivation are axioms. 

\paragraph{Proof search in Vampire.}
Given an input set of axioms $A$ and a conjecture $G$, \vampire{} searches for a refutation of $A \cup \{\neg G\}$, by using a preprocessing phase followed by a saturation phase.
In the preprocessing phase, \vampire{} generates a derivation from $A \cup \{\neg G\}$ such that each sink-node of the DAG\footnote{a sink-node is a node such that no edge emerges out of it.} is a clause. Then, \vampire{} enters the saturation phase, where it extends the existing derivation by applying its saturation algorithm using the sink-nodes from the preprocessing phase as the input clauses to saturation. The saturation phase of \vampire{} terminates in either of the following three cases:
(i) the empty clause $\bot$ is derived (hence, a proof  of $G$ was found), (ii) no more clauses are derived and the empty clause $\bot$ was not derived (hence, the input is saturated and $G$ is satisfiable),
or (iii) an a priory given time/memory limit on the \vampire{} run is reached (hence, it is unknown whether $G$ is satisfiable/valid).\\[-.75em]

% All state-of-the-art superposition-based theorem provers implement some version of the Given Clause Algorithm as their main saturation algorithm, and in particular either the Otter Algorithm [otter] or the Discount Algorithm [?]. In our context, the Discount algorithm should be seen as a simplified version of the Otter algorithm. We will therefore focus on the Otter algorithm.

% \begin{figure}[h]
% 	\centering
% 	\todoIn{}
% 	\caption{The Otter Saturation Algorithm}
% 	\label{alg-otter}
% \end{figure}

Saturation-based proving in \vampire{} is performed using the following high-level description of the saturation phase of \vampire{}.
The saturation algorithm divides the set of clauses from the proof space of \vampire{} into a set of $\mathit{Active}$ and $\mathit{Passive}$ clauses, and iteratively refines these sets using its superposition inference system: the $\mathit{Active}$ set keeps the clauses
between which all possible inferences have been performed, whereas the $\mathit{Passive}$ set stores the clauses which have not been added to $\mathit{Active}$ yet and are candidates for being used in future steps of the saturation algorithm.
During saturation, \vampire{} distinguishes between so-called {\it simplifying} and {\it generating inferences}. Intuitively,
simplifying inferences delete clauses from the search space and hence are crucial for  keeping the search space small.
A generating inference is a non-simplifying one, and hence adds new clauses to the search space. 
As such, at every iteration of the saturation algorithm, a new clause from $\mathit{Passive}$ is selected  and  added  to $\mathit{Active}$, after which all  generating
inferences between the selected clause and the clauses in $\mathit{Active}$ are applied. Conclusions of these inferences yield new clauses which are added to $\mathit{Passive}$ to be selected in future iterations of saturation.
Additionally at any step of the saturation algorithm, simplifying inferences and deletion of clauses are allowed.

\vspace*{-.5em}

\section{Analysis of Saturation Attempts of \vampire{}}\label{sec:analysis}

\vspace*{-.5em}

We now discuss how to efficiently analyze saturation attempts of \vampire{} in \satvis{}. \vspace*{-2em}

\paragraph{Analyzing saturation attempts.} To understand saturation (attempts),
we have to analyze the generating inferences performed during saturation (attempts).

On the one hand, we are interested in the \emph{useful} clauses:  that is, the derived and activated clauses that  are part of the proof we expect \vampire{} to find.
In particular, we check whether these clauses occur in $\mathit{Active}$. (i)  If this is the case for a given useful clause (or a simplified variant of it), we are done with processing  this useful clause and optionally check the derivation of that clause against the expected derivation. (ii) If not, we have to identify the reason why the clause was not added to $\mathit{Active}$, which can either be the case because (ii.a) the clause (or a simplified version of it) was never chosen from $\mathit{Passive}$ to be activated or (ii.b) the clause was not even added to $\mathit{Passive}$. In case (ii.a), we investigate why the clause was not activated. This involves checking which simplified version of the clause was added to $\mathit{Passive}$ and checking
the value of clause selection in \vampire{} on that clause.  
%the value assigned by the clause-selection-heuristic to that simplified clause against the maximal values assigned by that heuristic to any activated clause.
In case (ii.b), it is needed to understand why the clause was not added to $\mathit{Passive}$, that is,  why no generating inference between suitable premise clauses was performed. This could for instance be the case because one of the premises was not added to $\mathit{Active}$, in which case we recurse with the analysis on that premise, or because clause selection in \vampire{} prevented the inference.

On the other hand, we are interested in the \emph{useless} clauses:  that is, the clauses which were generated or even activated but are unrelated to the proof \vampire{} will find.
These clauses often slow down the proof search by several magnitudes. It is therefore crucial to limit their generation or at least their activation. To identify the useless clauses  that are activated, we need to analyze the set $\mathit{Active}$, whereas to identify the useless clauses, which are generated but never activated, we have to investigate the set $\mathit{Passive}$.

\begin{figure}[tb]
    \centering
    \begin{BVerbatim}[fontsize=\scriptsize]
...
[SA] passive: 160. v = a(l11(s(nl8)),$sum(i(main_end),1)) [superposition 70,118]
[SA] active: 163. i(main_end) != -1 [term algebras distinctness 162]
[SA] active: 92. ~'Sub'(X5,p(X4)) | 'Sub'(X5,X4) | zero = X4 [superposition 66,44]
[SA] new: 164. 'Sub'(p(p(X0)),X0) | zero = X0 | zero = p(X0) [resolution 92,94]
[SA] passive: 164. 'Sub'(p(p(X0)),X0) | zero = X0 | zero = p(X0) [resolution 92,94]
[SA] active: 132. v = a(l11(s(s(zero))),2) [superposition 70,124]
[SA] new: 165. v = a(l8(s(s(zero))),2) | i(l8(s(s(zero)))) = 2 [superposition 132,72]
[SA] new: 166. v = a(l8(s(s(zero))),2) | i(l8(s(s(zero)))) = 2 [superposition 72,132]
[SA] active: 90. s(X1) != X0 | p(X0) = X1 | zero = X0 [superposition 22,44]
[SA] new: 167. X0 != X1 | p(X0) = p(X1) | zero = X1 | zero = X0 [superposition 90,44]
[SA] new: 168. p(s(X0)) = X0 | zero = s(X0) [equality resolution 90]
[SA] new: 169. p(s(X0)) = X0 [term algebras distinctness 168]
... 
\end{BVerbatim}
    \caption{Screenshot of a saturation attempt of \vampire{}.
    \label{fig:screenshot-vampire-saturation}}\vspace*{-1em}
\end{figure}

\paragraph{Saturation output.}
%In the previous paragraph, we motivated why we are interested in the derivation and the sets $\mathit{Active}$ and $\mathit{Passive}$ induced by a saturation attempt.
We now discuss how \satvis{} reconstructs the clause sets $\mathit{Active}$ and $\mathit{Passive}$ from a \vampire{} saturation (attempt). 
\vampire{} is able to log a list of events, where each event is classified as either 
% providing sufficient information to reconstruct the derivation graph and the sets $\mathit{Active}$ and $\mathit{Passive}$. 
(i) new $C$ (ii) passive $C$ or (iii) active $C$, for a given clause $C$.
The list of events produced by \vampire{} satisfies the following properties:
(a) any clause is at most once newly created, added to $\mathit{Passive}$ and added to $\mathit{Active}$;
(b) if a clause is added to $\mathit{Passive}$, it was newly created in the same iteration,  and (c) if a clause is added to $\mathit{Active}$, it was newly created and added to $\mathit{Passive}$ at some point.
Figure~\ref{fig:screenshot-vampire-saturation} shows a part of the output logged by \vampire{} while performing a saturation attempt ({\tt SA}).

Starting from an empty derivation and two empty sets, the derivation graph and the sets $\mathit{Active}$ and $\mathit{Passive}$ corresponding to a given saturation attempt of \vampire{} are computed in \satvis{} by traversing the list of events produced by \vampire{} and iteratively changing the derivation and the
sets $\mathit{Active}$ and $\mathit{Passive}$, as follows: 
\begin{enumerate}[label=(\roman*)]
    \item new $C$: add the new node $C$ to the derivation and construct the edges $(C_i,C)$, for any premise $C_i$ of the inference deriving $C$. The sets $\mathit{Active}$ or $\mathit{Passive}$ remain unchanged; 
    \item passive $C$: add the node $C$ to $\mathit{Passive}$. The derivation and $\mathit{Active}$ remain unchanged; 
    \item active $C$: remove the node $C$ from $\mathit{Passive}$ and add it to $\mathit{Active}$. The derivation remains unchanged.
    % \item “[SA] delete $C$” does not change the derivation. If $C$ is contained in $\mathit{Passive}$ or $\mathit{Active}$, $C$ is removed from that set and the other set stays the same. Otherwise both sets stay the same.
\end{enumerate}

\vspace*{-1em}
\paragraph{Interactive Visualization.} 
The large number of inferences during saturation in \vampire{} makes the direct  analysis of saturation attempts of \vampire{}
impossible within  a reasonable amount of time.
In order to overcome this problem, in \satvis{} we \emph{interactively} 
visualize the derivation graph of the \vampire{} saturation. The graph-based visualization of \satvis{} brings the following benefits: 

$\bullet$ Navigating through the graph visualization of a \vampire{} derivation is easier for users
rather than working with the \vampire{} derivation
encoded as a list of hyper-edges. In particular,
both (i) navigating to the premises of a selected node/clause and (ii) searching for inferences having a selected node/clause as premise is performed fast in \satvis{}.

$\bullet$ \satvis{} visualizes only the  nodes/clauses that are part of a derivation of an activated clause, and in this way ignores uninteresting inferences.

$\bullet$ \satvis{} merges the preprocessing inferences, such that each clause resulting from preprocessing has as direct premise the input formula it is derived from.\\[-.75em]

Yet, a straightforward graph-based visualization of \vampire{} saturations in \satvis{} would bring the following practical limitations on using \satvis{}:

(i) displaying  additional meta-information on graph nodes, such as the inference rule used to derive a node, is computationally very expensive, due to the large number of inferences used during saturation;

(ii) manual search for particular/already processed nodes in relatively large derivations would take too much time;

(iii) subderivations are often interleaved with other subderivations due to an imperfect automatic layout of the graph.\\[-.75em]

\satvis{} addresses the above challenges using its following interactive features: 

\begin{itemize}
    \item \satvis{} displays meta-information only for a selected node/clause; 
    \item \satvis{} supports different ways to locate and select clauses, such as full-text search, search for direct children and premises of the currently selected clauses, and search for clauses whose derivation contains all currently selected nodes; 
    \item \satvis{} supports transformations/fragmentations of derivations.
      In particular, it is possible  to restrict and visualize the derivation containing only the clauses that form the derivation of a selected clause,
      or visualize only clauses whose derivation contains a selected clause.
    \item \satvis{} allows to (permanently) highlight one or more clauses in the derivation.
\end{itemize}

\begin{figure}[tb]
    \centering
    \includegraphics[width=0.6\textwidth]{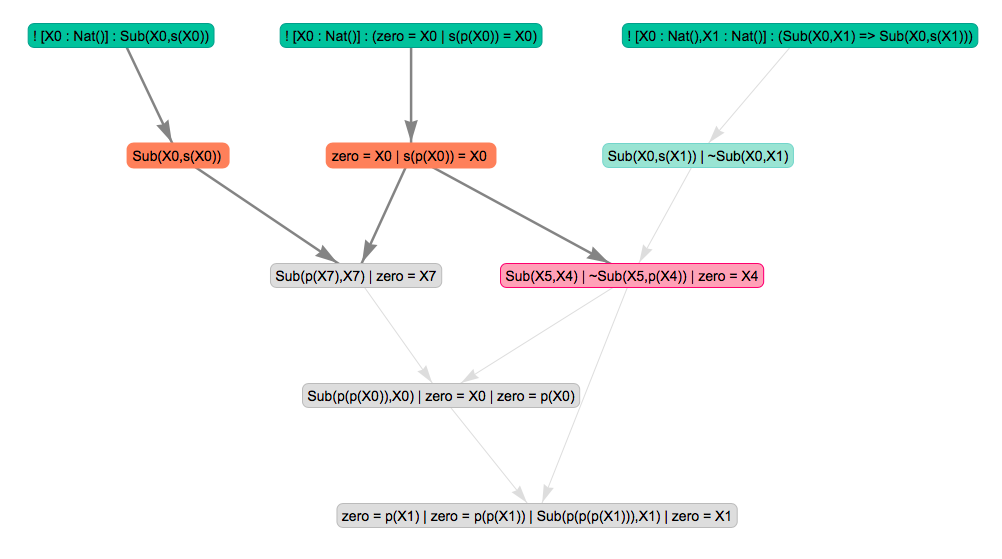}
    \includegraphics[width=0.21\textwidth]{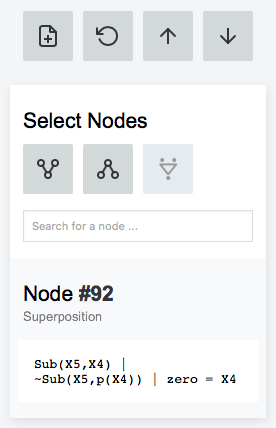}
    \caption{Screenshot of \satvis{} showing visualized derivation and interaction menu\label{fig:visualize}.}\vspace*{-1.5em}
    \label{}
\end{figure}

Figure~\ref{fig:visualize} illustrates some of the above feature of \satvis{}, using output from \vampire{} similar to Figure~\ref{fig:screenshot-vampire-saturation} as input to \satvis{}.

\vspace*{-.5em}

\section{Implementation of \satvis{} 1.0}\label{sec:implementation}

\vspace*{-.5em}

We implemented \satvis{} as a web application, allowing \satvis{} to be easily used on any platform.
Written in Python3, \satvis{} contains about 2,200 lines of code. For the generation of graph layouts, we rely on {\tt pygraphviz}\footnote{\url{https://pygraphviz.github.io}}, whereas graph/derivation  visualizations are created with {\tt vis.js}\footnote{\url{https://visjs.org/}}. We experimented with \satvis{} on the verification examples of~\cite{rapidArxiv},
using an Intel Core i5 3.1Ghz machine with 16 GB of RAM, allowing us to refine and successfully generate \vampire{} proofs for non-interference and information-flow examples  of~\cite{rapidArxiv}.

\vspace*{-1em}

\paragraph{\satvis{} workflow.}
\satvis{} takes as input a text file containing the output of a \vampire{}  saturation attempt. An example of a partial input to \satvis{} is given in Figure~\ref{fig:screenshot-vampire-saturation}.
\satvis{} then generates a DAG representing the derivation of the considered \vampire{} saturation output, as presented in Section~\ref{sec:analysis} and discussed later. 
Next, \satvis{} generates the graph layout of for the generated DAG, enriched with configured style information.
Finally, \satvis{} renders and visualizes the \vampire{} derivation corresponding to its input, and
allows interactive visualisations of its output, as discussed in Section~\ref{sec:analysis} and detailed below. 
%Additional elements of the user interface allow to interact with the graph. %The workflow applied in the implementation of \satvis{} is displayed in Fig. \ref{workflow}.

%\begin{figure}[]
 %   \centering
  %  \begin{tikzpicture}
  %    \node (input) at (0,0) {Input};
  %    \node (parser) at (2,0) {Parser};
  %    \node (preproc) at (4.2,0) {Preprocessing};
  %    \node (graphviz) at (6.7,0) {pygraphviz};
  %    \node (visjs) at (9,0) {vis.js};
  %    \node (static) at (4.5, -0.8) {\emph{\small Visualization}};
   %   \node (interactive) at (9, -0.8) {\emph{\small Interaction}};
   %   \draw[->] (input) to (parser);
   %   \draw[->] (parser) to (preproc);
    %  \draw[->] (preproc) to (graphviz);
    %  \draw[->] (graphviz) to (visjs);
    %  \draw (1,0.5) -- (8,0.5) -- (8, -0.5) -- (1, -0.5) -- (1,0.5);
     % \draw (8.2,0.5) -- (9.8,0.5) -- (9.8, -0.5) -- (8.2, -0.5) -- (8.2,0.5);
   % \end{tikzpicture}
   % \caption{Implementation Workflow of \satvis{}}
   % \label{workflow}
%\end{figure}

%
%\subsection{Design}
%\satvis{} comprises three major parts: (i) the parsing and analyzing of \vampire{} output, (ii) the visualization and (iii) interactions with the generated graph.

%\paragraph{Input}
%\satvis{} takes as input a \vampire{} logfile.

\vspace*{-.5em}
\paragraph{DAG generation of saturation outputs.}
\satvis{} parses its input line by line using regex pattern matching in order to generate the nodes of the graph.
Next, \satvis{} uses a post order traversal algorithm to sanitize nodes and remove redundant ones.
The result is then passed to {\tt pygraphviz} to generate a graph layout. While {\tt pygraphviz} finds layouts for thousands of nodes within less than three seconds, 
we would like to improve the scalability of the tool further.
% we note that the preprocessing step does not scale well. 

It would be beneficial to preprocess and render nodes incrementally, while ensuring stable layouts for \satvis{} graph transformations. We leave this engineering task for future work. 

%The number of nodes in a typical \satvis{} graph can reach the five figure mark. If we displayed the graph as a static image,
%finding relevant nodes and inferences could still be tedious and time-consuming. To avoid this problem, we add interactive features to \satvis{}.

\vspace*{-.5em}
\paragraph{Interactive visualization}
The interactive features of \satvis{} support (i) various node searching mechanisms, (ii) graph transformations, and (iii) the display of meta-information about a specific node.
We can efficiently search for nodes by (partial) clause, find parents or children of a node, and find common consequences of a number of nodes.
Graph transformations in \satvis{} allow to only render a certain subset of nodes from the \satvis{} DAG, for example, displaying only transitive parents or children of a certain node.
% \satvis{} can be extended to display more meta-information, such as inference rules used to derive a certain node or the number of children of a selected node.

 \vspace*{-.5em}

\section{Conclusion}
\vspace*{-.5em}

We described the \satvis{} tool for interactively visualizing proofs and proof
attempts of the first-order theorem prover \vampire{}.
Our work analyses proof search in \vampire{} and reconstructs
first-order derivations corresponding to \vampire{} proofs/proof
attempts. The interactive features of \satvis{} ease
the task of understanding both successful and failing proof attempts
in \vampire{} and hence can be used to further develop and use 
\vampire{} both by experts and non-experts in first-order theorem
proving.

\vspace*{0.5em}

\smallskip
\noindent\textbf{Acknowledgements.}
This work was funded by the ERC Starting Grant 2014
SYMCAR 639270, the ERC Proof of Concept Grant 2018
SYMELS 842066, the Wallenberg Academy Fellowship 2014
TheProSE and the Austrian FWF project W1255-N23.

\bibliographystyle{abbrv}
\bibliography{bib.bib}

\begin{thebibliography}{10}

\bibitem{CVC4}
C.~Barrett, C.~L. Conway, M.~Deters, L.~Hadarean, D.~Jovanovi{\'c}, T.~King,
  A.~Reynolds, and C.~Tinelli.
\newblock {CVC4}.
\newblock In {\em CAV}, pages 171--177, 2011.

\bibitem{barrett2017smtlib}
C.~Barrett, P.~Fontaine, and C.~Tinelli.
\newblock The {SMT-LIB} standard: Version 2.6.
\newblock Technical report, Department of Computer Science, The University of
  Iowa, 2017.

\bibitem{rapidArxiv}
G.~{Barthe}, R.~{Eilers}, P.~{Georgiou}, B.~{Gleiss}, L.~{Kovacs}, and
  M.~{Maffei}.
\newblock {Verifying Relational Properties using Trace Logic}.
\newblock In {\em FMCAD}, 2019.
\newblock To appear.

\bibitem{BYRNES200923}
J.~Byrnes, M.~Buchanan, M.~Ernst, P.~Miller, C.~Roberts, and R.~Keller.
\newblock Visualizing proof search for theorem prover development.
\newblock {\em ENTCS}, 226:23 -- 38, 2009.

\bibitem{Z3}
L.~De~Moura and N.~Bj{\o}rner.
\newblock {Z3}: An efficient {SMT} solver.
\newblock In {\em TACAS}, pages 337--340, 2008.

\bibitem{Gansner00anopen}
E.~R. Gansner and S.~C. North.
\newblock {An Open Graph Visualization System and its Applications to Software
  Engineering}.
\newblock {\em Software- Practice and Experience}, 30(11):1203--1233, 2000.

\bibitem{FOOL18}
E.~Kotelnikov, L.~Kov{\'{a}}cs, and A.~Voronkov.
\newblock {A FOOLish Encoding of the Next State Relations of Imperative
  Programs}.
\newblock In {\em IJCAR}, pages 405--421, 2018.

\bibitem{kovacs2017coming}
L.~Kov{\'a}cs, S.~Robillard, and A.~Voronkov.
\newblock Coming to terms with quantified reasoning.
\newblock In {\em POPL}, pages 260--270. ACM, 2017.

\bibitem{kovacs2013first}
L.~Kov{\'a}cs and A.~Voronkov.
\newblock {First-Order Theorem Proving and Vampire}.
\newblock In {\em CAV}, pages 1--35, 2013.

\bibitem{DBLP:journals/corr/LibalRR14}
T.~Libal, M.~Riener, and M.~Rukhaia.
\newblock {Advanced Proof Viewing in ProofTool}.
\newblock In {\em UITP}, pages 35--47, 2014.

\bibitem{Sup01}
R.~Nieuwenhuis and A.~Rubio.
\newblock {Paramodulation-Based Theorem Proving}.
\newblock In {\em Handbook of Automated Reasoning}, pages 371--443. 2001.

\bibitem{z3-axiom-profiler}
F.~Rothenberger.
\newblock Integration and analysis of alternative smt solvers for software
  verification.
\newblock Master's thesis, ETH Zurich, Zürich, 2016.
\newblock Masterarbeit. ETH Zürich. 2016.

\bibitem{ESchulz}
S.~Schulz.
\newblock {E - a Brainiac Theorem Prover}.
\newblock {\em {AI} Communications}, 15(2-3):111--126, 2002.

\bibitem{Sut07-CSR}
G.~Sutcliffe.
\newblock {TPTP, TSTP, CASC, etc.}
\newblock In {\em {CSR}}, pages 7--23, 2007.

\bibitem{DRAT}
N.~Wetzler, M.~J.~H. Heule, and W.~A. Hunt.
\newblock Drat-trim: Efficient checking and trimming using expressive clausal
  proofs.
\newblock In {\em SAT}, pages 422--429, 2014.

\end{thebibliography}

\end{document}